\documentclass[showpacs,aps,prb,twocolumn]{revtex4}
\usepackage{amsmath}
\usepackage{amsfonts}
\usepackage{amssymb}
\usepackage{amsthm}
\usepackage{graphicx}
\usepackage{color}
\begin{document}


\title{Coherent Quantum Dynamics: What Fluctuations Can Tell}
\author{John Schliemann}
\affiliation{ Institute for Theoretical Physics, University of
Regensburg, D-93040 Regensburg, Germany}
\date{May 2015}

\begin{abstract}
Coherent states provide a natural connection of quantum systems to their
classical limit and are employed in various fields of physics. Here we derive
general systematic expansions, with respect to quantum parameters, of
expectation values of products of arbitrary operators within both
oscillator coherent states and SU(2) coherent states. In particular, we 
generally prove that the energy fluctuations of an arbitrary Hamiltonian
are in leading order entirely due to the time dependence of the classical
variables. These results add to the list of wellknown properties of
coherent states and are applied here to the Lipkin-Meshkov-Glick model, the
Dicke model, and to coherent intertwiners in spin networks as considered in
Loop Quantum Gravity.   
\end{abstract}
%
%
\maketitle

\section{introduction}

Coherent states are at the heart of semiclassical descriptions of generic
quantum systems and have proven to be a versatile tool in a multitude
of physical problems. In the general literature 
\cite{Klauder85,Perelomov86,Zhang90,Barnett97,Gazeau09}, mainly two types of
coherent states are typically distinguished: The first type, the coherent
states of the harmonic oscillator, was already investigated by Schr\"odinger
\cite{Schrodinger26} shortly after the birth of quantum mechanics, while
SU(2) coherent states in the Hilbert space of a spin of general length $S$
were added in the early 1970s \cite{Radcliffe71,Arecchi72}.

Both types of coherent states share a list of wellknown properties which
constitute the basis for their prominent role in semiclassics: (i) The coherent
states can be generated by a unitary transformation from an appropriate 
reference state. In the oscillatory case this state is the ground state
of an harmonic, while for spins one uses the heighest-weight
state in some arbitrary basis. As a result, the coherent states are (ii)
(over-)complete, 
(iii) eigenstates of simple operators generic to the system, and (iv)
they have minimum uncertainty products with respect to an obvious choice of
variables. Moreover, (v) coherent states show a coherent time evolution
perfectly mimicking the classical limit under appropriate Hamiltonians.
For oscillator coherent states such a Hamiltonian is the one of the harmonic
oscillator itself, and for the spin case the Zeeman Hamiltonian 
(coupling the spin to an external magnetic field) plays an analogous role.

In the present work we argue that one can extend the above list by general
statements about correlations and fluctuations within coherent states.
Specifically we consider the coherent expectation value of a product of two
arbitrary operators. For such expectation values we derive systematic
expansions in the quantum parameters $\hbar$ or $1/S$ which involve only
coherent expectation values of single operators and their commutators
with the system variables. These expansions are a versatile tools for the
study of the semiclassical regime of generic quantum systems.
As an important finding, 
the energy fluctuations of an arbitrary Hamiltonian are generally proven to be
in leading order entirely due to the time dependence of the classical
variables. These results add to the above list of properties of
coherent states. Reflecting the widespread use of the latter objects, we 
apply our findings to the Lipkin-Meshkov-Glick model originating from
nuclear physics \cite{Lipkin65}, to the Dicke model describing superradiance 
in quantum optics \cite{Dicke54}, and to coherent intertwiners of spin 
networks occurring in the loop approach to quantum gravity
\cite{Livine07,Rovelli14}.

This paper is organized as follows: In section \ref{revcostate}
we review and summarize important properties of oscillator coherent states and
SU(2) coherent states. The announced results on the coherent expectation values
of arbitrary operator products are derived in section \ref{corr} and discussed
there on a general footing. Some technical details of the calculations are
deferred to appendix \ref{N>1}.
Section \ref{lmg} contains the application of our general findings to
the Lipkin-Meshkov-Glick model, and the Dicke model is treated in
section \ref{dicke}. In section \ref{intertw} we turn to the study of
coherent intertwiners of spin networks investigated in the covariant approach
advocated by Loop Quantum Gravity. Here we derive semiclassical corrections
to expectation values in terms of universal expansion coefficients 
depending only on the network geometry.
We close with a summary and an outlook in section \ref{concl}.

\section{Coherent States}

\label{revcostate}

We now briefly review, using standard notation, distinctive properties
of oscillator coherent states and SU(2) coherent states.

\subsection{Coherent Oscillator States}
\label{revosc}

The harmonic oscillator is described by
\begin{equation}
{\cal H}_{h}=\frac{1}{2}\left(p^2+\omega^2q^2\right)
=\hbar\omega\left(a^{+}a+\frac{1}{2}\right)
\label{hH}
\end{equation}
with
\begin{equation}
a=\frac{1}{\sqrt{2}}\left(\sqrt{\frac{\omega}{\hbar}}q
+\frac{i}{\sqrt{\hbar\omega}}p\right)\quad,\quad 
a^+=\left(a\right)^+
\label{aa}
\end{equation}
fulfilling
\begin{equation}
\left[p,q\right]=\frac{\hbar}{i}
\quad\Leftrightarrow\quad\left[a,a^{+}\right]=1\,.
\label{cr}
\end{equation}
The system has an equidistant spectrum labelled by $n\in\{0,1,2,\dots\}$,
\begin{equation}
{\cal H}_{h}|n\rangle=\hbar\omega\left(n+\frac{1}{2}\right)|n\rangle\,.
\end{equation} 
Coherent states of the harmonic oscillator are eigenstates of the lowering
operator $a$ with complex eigenvalues $\alpha$,
\begin{equation}
a|\alpha\rangle=\alpha|\alpha\rangle\,.
\label{defcs}
\end{equation}
They are generated from the ground state via
\begin{eqnarray}
|\alpha\rangle & = & \exp\left(\alpha a^{+}-\alpha^{\ast}a\right)|0\rangle
\label{cs1}\\
& = & \exp\left(-\frac{1}{2}|\alpha|^{2}\right)
\sum_{n=0}^{\infty}\frac{\alpha^{n}}{\sqrt{n!}}|n\rangle\,.
\label{cs2}
\end{eqnarray}
The parameter $\alpha$ is naturally decomposed into its real and imaginary
part as
\begin{equation}
\alpha=\frac{1}{\sqrt{2}}\left(\sqrt{\frac{\omega}{\hbar}}\xi
+\frac{i}{\sqrt{\hbar\omega}}\pi\right)\,.
\label{alphadecomp}
\end{equation}
Denoting an expectation value within a coherent state (\ref{cs1}) by
$\langle\cdot\rangle$ it holds
\begin{equation}
\langle q\rangle=\xi\quad,\quad\langle p\rangle=\pi\,.
\label{exval1}
\end{equation}
Coherent states maintain their shape in the time evolution of the harmonic
oscillator,
\begin{equation}
e^{-\frac{i}{\hbar}{\cal H}_ht}|\alpha\rangle
=e^{-\frac{i}{2}\omega t}|\alpha e^{-i\omega t}\rangle\,,
\end{equation}
and the time dependence of the expectation values (\ref{exval1}) follows
exactly the classical motion of the harmonic oscillator. This fact justifies
the term `coherent states' and relies on the equidistance of the spectrum.
The latter property is shared by a quantum spin of arbitrary length in a 
magnetic field and leads there to a coherent Larmor precession, as we will
discuss in section \ref{revSU2}.

Moreover, coherent states minimize uncertainty 
products,
\begin{equation}
\Delta p\Delta q=\hbar/ 2\,
\label{uncertosc}
\end{equation}
and fulfill an (over-)completeness relation,
\begin{equation}
\frac{1}{\pi}\int d^{2}\alpha|\alpha\rangle\langle\alpha|={\bf 1}\,.
\label{complosc}
\end{equation} 

\subsection{{\rm SU(2)} coherent states}
\label{revSU2}

In the Hilbert space of a spin of length $S$ an SU(2) (or spin) coherent 
state $|\vartheta,\varphi\rangle$ is defined by the equation
\begin{equation}
\vec s\cdot{\vec S}|\vartheta,\varphi\rangle
=\hbar S|\vartheta,\varphi\rangle
\label{defsu2}
\end{equation}
for the direction $\vec s=
(\sin\vartheta\cos\varphi,\sin\vartheta\sin\varphi,\cos\vartheta)$.
For generic systems expectation values within these states provide a natural
approach to the classical limit given by
$\hbar\rightarrow 0$, $S\rightarrow\infty$
while $\hbar S$ is kept constant.
 
Introducing the usual basis of eigenstates of $S^z$ 
($S^z |m\rangle=\hbar m|m\rangle$) coherent states can be
generated from $|S\rangle$ by a unitary rotation,
\begin{eqnarray} 
|\vartheta,\varphi\rangle & = & U(\vartheta,\varphi)\,|S\rangle
\label{su2cos1}\\
& = & \frac{1}{(1+|z|^2)^S}e^{zS^-}|S\rangle
\label{su2cos2}
\end{eqnarray}
with \cite{Perelomov86,Barnett97}
\begin{eqnarray}
U(\vartheta,\varphi) & = & \exp\left(\frac{i}{\hbar}\vartheta
\left(\sin\varphi S^x-\cos\varphi S^y\right)\right)
\label{su2U1}\\
& = & e^{zS^-/ \hbar}e^{\eta S^z/ \hbar}e^{-\bar zS^+/ \hbar}
\label{su2U2}\\
& = & e^{-\bar zS^+/ \hbar}e^{-\eta S^z/ \hbar}e^{zS^-/ \hbar}
\label{su2U3}
\end{eqnarray}
and 
\begin{equation}
z(\vartheta,\varphi)=\tan\frac{\vartheta}{2}e^{i\varphi}\quad,\quad
\eta(\vartheta)=2\ln\cos\frac{\vartheta}{2}\,.
\end{equation}
Expanded in the above basis SU(2) coherent states read
\begin{eqnarray} 
|\vartheta,\varphi\rangle & = & \frac{1}{(1+|z|^2)^S} 
\sum_{m=-S}^S{2S\choose S+m}^{\frac{1}{2}}z^m|m\rangle\\
& = & \sum_{m=-S}^S\Biggl[{2S\choose S+m}^{\frac{1}{2}}
\left(\cos\left(\frac{\vartheta}{2}\right)\right)^{S+m}\nonumber\\
& & \qquad\cdot\left(\sin\left(\frac{\vartheta}{2}\right)\right)^{S-m}
e^{\imath\varphi(s+m)}\,|m\rangle\Biggr]\,.
\end{eqnarray}
The analog of the harmonic oscillator for SU(2) coherent states is the
Zeeman Hamiltonian
\begin{equation}
{\cal H}_z=-\vec S\cdot\vec h
\label{zeeman}
\end{equation}
coupling the spin to a magnetic field $\vec h$. The spectrum consists of
$2S+1$ equidistant energy levels, and the corresponding time evolution
of SU(2) coherent states is a coherent Larmor precession, which is most easily
seen when putting, without loss of 
generality, the field direction along the $z$-axis, 
\begin{equation}
e^{-\frac{i}{\hbar}{\cal H}_zt}|\vartheta,\varphi\rangle
=e^{-i\varphi Sht}|\vartheta,\varphi+ht\rangle\,.
\end{equation}
The latter finding is completely analogous to the harmonic oscillator having a
semi-infinite equidistant spectrum.

As further standard properties shared with coherent oscillator states,
SU(2) coherent states have a minimum uncertainty product
\begin{equation}
\Delta(\vec e_{1}\cdot{\vec S}\,)\,\Delta(\vec e_{2}\cdot{\vec S}\,)=
\frac{\hbar^2S}{2}
\end{equation}
with $\vec e_{1}$,$\vec e_{2}$,$\vec s$ being an orthonormal system, and 
their (over-) completeness can be expressed as
\begin{eqnarray}
{\bf 1} & = & 
\frac{2S+1}{4\pi}\int_0^{2\pi}d\varphi\int_0^{\pi}d\vartheta\sin\vartheta
|\vartheta,\varphi\rangle\langle\vartheta,\varphi|
\label{complsu21}\\
& = & \frac{2S+1}{\pi}\int\frac{d^2z}{(1+|z|^2)^2}|z\rangle\langle z|
\label{complsu22}\\
& = & \frac{2S+1}{\pi}\int d^2z\frac{e^{zS^-/ \hbar}|S\rangle
\langle S|e^{\bar zS^+/ \hbar}}{(1+|z|^2)^{2(S+1)}}\,,
\label{complsu23}
\end{eqnarray}
where $|z\rangle=|\vartheta,\varphi\rangle$. Further below it will
be useful to change reference state $|S\rangle$ in Eq.~(\ref{complsu23})
to an arbitrary SU(2) coherent state by applying the unitary transformation
given in Eqs.~(\ref{su2U1})-(\ref{su2U3}):
\begin{eqnarray}
{\bf 1} & = & \frac{2S+1}{\pi}U\int d^2w\frac{e^{wS^-/ \hbar}|S\rangle
\langle S|e^{\bar wS^+/ \hbar}}{(1+|w|^2)^{2(S+1)}}U^+\nonumber\\
& = & \frac{2S+1}{\pi}\int d^2w\frac{e^{w\tilde S^-/ \hbar}|z\rangle
\langle z|e^{\bar w\tilde S^+/ \hbar}}{(1+|w|^2)^{2(S+1)}}
\label{complsu24}
\end{eqnarray}
with $\vec{\tilde S}=U\vec SU^+$.

\section{Correlations}
\label{corr}

We now derive general theorems for the expectation values of operator products
within coherent states.

\subsection{Oscillatory Systems}

\subsubsection{General Correlation Functions}

Let $A$, $B$ be two operators being functions of the two canonical operators
$p$, $q$ (or, equivalently, $a$, $a^+$). Using the completeness relation
(\ref{complosc}) the expectation value of $AB$ within a coherent oscillator 
state can be formulated as
\begin{eqnarray}
\langle\alpha|AB|\alpha\rangle & = & \frac{1}{\pi}\int d^2\beta e^{-|\beta|^2}
\langle 0|U_{\alpha}^+AU_{\alpha}U_{\alpha}^+e^{\beta a^+}|0\rangle\nonumber\\
& & \qquad\qquad
\cdot\langle 0|e^{\bar\beta a}U_{\alpha}U_{\alpha}^+BU_{\alpha}|0\rangle\,,
\end{eqnarray}
where $U_{\alpha}$ is the unitary operator on the r.h.s. of Eq.~(\ref{cs1}), and
\begin{equation}
U_{\alpha}^+e^{\beta a^+}|0\rangle=e^{-\frac{1}{2}|\alpha|^2+\bar\alpha\beta}
e^{(\beta-\alpha)a^+}|0\rangle
\end{equation}
such that
\begin{eqnarray}
\langle\alpha|AB|\alpha\rangle & = & \frac{1}{\pi}\int d^2\beta 
e^{-|\beta-\alpha|^2}
\langle 0|U_{\alpha}^+AU_{\alpha}e^{(\beta-\alpha)a^+}|0\rangle\nonumber\\
& & \qquad\qquad
\cdot\langle 0|e^{(\bar\beta-\bar\alpha)a}U_{\alpha}^+BU_{\alpha}|0\rangle\nonumber\\
& = & \frac{1}{\pi}\int d^2\beta 
e^{-|\beta|^2}
\langle 0|e^{-\beta a^+}U_{\alpha}^+AU_{\alpha}e^{\beta a^+}|0\rangle\nonumber\\
& & \qquad\qquad
\cdot\langle 0|e^{\bar\beta a}U_{\alpha}^+BU_{\alpha}e^{-\bar\beta a}|0\rangle\,,
\end{eqnarray}
where we have shifted the integration variable and used
$e^{-\bar\beta a}|0\rangle=|0\rangle$. The remaining operator products can be
expanded into series of iterated commutators according to
\begin{equation}
e^XYe^{-X}=\sum_{n=0}^{\infty}\frac{1}{n!}\left[X,Y\right]_n
\label{itcom}
\end{equation}
with $[X,Y]_0=Y$ and $[X,Y]_n=[X,[X,Y]_{n-1}]$. Upon performing the integration 
the two infinite series shrink to a single one yielding
\begin{eqnarray}
\langle\alpha|AB|\alpha\rangle & = & \sum_{n=0}^{\infty}\frac{1}{n!}
\langle 0|\left[-a^+,U_{\alpha}^+AU_{\alpha}\right]_n|0\rangle\nonumber\\
& & \qquad\qquad
\cdot\langle 0|\left[a,U_{\alpha}^+BU_{\alpha}\right]_n|0\rangle
\label{oscexp1a}\\
& = & \sum_{n=0}^{\infty}\frac{1}{n!}
\langle\alpha|\left[iU_{\alpha}a^+U_{\alpha}^+,A\right]_n|\alpha\rangle\nonumber\\
& & \qquad\qquad
\cdot\langle\alpha|\left[iU_{\alpha}aU_{\alpha}^+,B\right]_n
|\alpha\rangle\nonumber\\
& = & \sum_{n=0}^{\infty}\frac{1}{n!}
\langle\alpha|\left[ia^+,A\right]_n|\alpha\rangle\nonumber\\
& & \qquad\qquad
\cdot\langle\alpha|\left[ia,B\right]_n|\alpha\rangle\,.
\label{oscexp1}
\end{eqnarray}
In the last step we took into account that $U_{\alpha}a^+U_{\alpha}^+$ and
$a^+$ differ just by a constant which commutes with any operator. Thus,
we have arrived at an expression for the expectation value of product
of two operators within coherent states in terms of a sum over products of
such coherent-state expectation values which involve only one of the operators. 
An alternative form of the above expansions can be given via 
Eq.~(\ref{oscexp1a}) as
\begin{equation}
\langle\alpha|AB|\alpha\rangle=\sum_{n=0}^{\infty}
\langle 0|U_{\alpha}^+AU_{\alpha}|n\rangle\
\langle n|U_{\alpha}^+BU_{\alpha}|0\rangle\,,
\label{oscexp1b}
\end{equation}
which of course just expresses the completeness of the states $|n\rangle$
and provides an alternative way to derive Eq.~(\ref{oscexp1}).

Moreover, using the definition (\ref{aa}) Eq.~(\ref{oscexp1}) can be 
rewritten as
\begin{eqnarray}
\langle\alpha|AB|\alpha\rangle & = & \sum_{n=0}^{\infty}\frac{\hbar^n}{n!2^n}
\left\langle\alpha\left|\left[
\frac{i}{\hbar}\left(\sqrt{\omega}q-i\frac{p}{\sqrt{\omega}}\right),
A\right]_n\right|\alpha\right\rangle\nonumber\\
& & \cdot\left\langle\alpha\left|\left[
\frac{i}{\hbar}\left(\sqrt{\omega}q+i\frac{p}{\sqrt{\omega}}\right),
B\right]_n\right|\alpha\right\rangle\,.
\label{oscexp2}
\end{eqnarray}
Since each commutation of $p$, $q$ with $A$ or $B$ yields a factor of $\hbar$
all expectation value on the above r.h.s. are of the same
order in $\hbar$. Thus, Eq.~(\ref{oscexp2}) is indeed a systematic expansion 
in $\hbar$ of the coherent-state expectation value of an arbitrary product of 
two operators. The zeroth order equals the classical result,
and for a general correlation function one has the semiclassical
expansion
\begin{eqnarray}
C_{AB} & := & \langle\alpha|AB|\alpha\rangle
-\langle\alpha|A|\alpha\rangle\langle\alpha|B|\alpha\rangle\nonumber\\
& = & \sum_{n=1}^{\infty}\frac{\hbar^n}{n!2^n}
\left\langle\alpha\left|\left[
\frac{i}{\hbar}\left(\sqrt{\omega}q-i\frac{p}{\sqrt{\omega}}\right),
A\right]_n\right|\alpha\right\rangle\nonumber\\
& & \cdot\left\langle\alpha\left|\left[
\frac{i}{\hbar}\left(\sqrt{\omega}q+i\frac{p}{\sqrt{\omega}}\right),
B\right]_n\right|\alpha\right\rangle\,.
\label{oscexp3}
\end{eqnarray}
Choosing $A=B$ we obtain a general expression for the variance of an hermitian
operator $A$,
\begin{eqnarray}
\left(\Delta A\right)^2 & = & \sum_{n=1}^{\infty}\frac{\hbar^n}{n!2^n}
\left|\left\langle\alpha\left|\left[
\frac{i}{\hbar}\left(\sqrt{\omega}q-i\frac{p}{\sqrt{\omega}}\right),
A\right]_n\right|\alpha\right\rangle\right|^2\,,\nonumber\\
\label{oscexp4}
\end{eqnarray}
where each term in the semiclassical expansion is nonnegative. 

\subsubsection{Energy Fluctuations}

Considering $A={\cal H}$ as an Hamiltonian, the corresponding energy 
fluctuation reads in leading order in $\hbar$
\begin{eqnarray}
\left(\Delta{\cal H}\right)^2 & = & \frac{\hbar}{2}\left(
\left\langle\frac{i}{\hbar}\left[\sqrt{\omega}q,{\cal H}\right]\right\rangle^2
+\left\langle\frac{i}{\hbar}\left[\frac{p}{\sqrt{\omega}},{\cal H}\right]
\right\rangle^2\right)\nonumber\\
& & \qquad+{\cal O}\left(\hbar^2\right)
\label{oscergfluc1}\\
& = & \frac{\hbar}{2}\left(\omega\left\langle\partial_tq\right\rangle^2
+\frac{\left\langle\partial_tp\right\rangle^2}{\omega}\right)\,,
\label{oscergfluc2}
\end{eqnarray}
where we have replaced, according to the Heisenberg equations of motion, the
commutators with time derivatives. Indeed, if the system is prepared at some
initial time $t=t_i$ in a coherent state we have (cf. Eq.~(\ref{alphadecomp}))
\begin{equation}
\left\langle\partial_tq\right\rangle=\partial_t\xi\quad,\quad
\left\langle\partial_tp\right\rangle=\partial_t\pi
\end{equation}
and 
\begin{equation}
\left(\Delta{\cal H}\right)^2
=\frac{\hbar}{2}\left(\omega\left(\partial_t\xi\right)^2
+\frac{\left(\partial_t\pi\right)^2}{\omega}\right)
+{\cal O}\left(\hbar^2\right)
\label{oscergfluc3}
\end{equation}
at $t=t_i$. In the subsequent time evolution governed by the Hamiltonian
${\cal H}$ the state of the system will, for not too large times, approximately
be coherent with time-dependent parameters $\xi(t)$, $\pi(t)$ playing the
approximate role of classical Hamiltonian variables. Thus, in this
semiclassical regime the fact that a coherent state has a finite energy
variance, i.e. it is not an eigenstate of the Hamiltonian, is in leading
order in $\hbar$ just expressed by the fact that the classical Hamiltonian
variables have a nontrivial time dependence, i.e. the system is moving.
This result complements the historical Ehrenfest theorem stating that
expectation values of observables follow the classical equations.

Relations of the type (\ref{oscergfluc2}),(\ref{oscergfluc3}) were already
found in Ref.~\cite{Schliemann99} on the example of specific Hamiltonians.
The results here are derived for arbitrary systems and are based
on the very general expansions (\ref{oscexp3}),(\ref{oscexp4}) for
correlation functions and fluctuations.

The fact that the system will in its time evolution in general not strictly
remain in a coherent state, i.e. decoherence occurs, 
is reflected by the higher contributions to
the energy variance. Indeed, for a harmonic oscillator (\ref{hH}) 
the time evolution is strictly coherent and we have as an identity
\begin{equation}
\left(\Delta{\cal H}_h\right)^2
\equiv\frac{\hbar}{2}\left(\omega\left(\partial_t\xi\right)^2
+\frac{\left(\partial_t\pi\right)^2}{\omega}\right)
\label{oscergfluc4}
\end{equation}
for all times $t\geq t_i$ and without any higher correction.

Finally, it is straightforward to extend the above results for general
operator products to the
case of $N>1$ degrees of freedom; details are sketched in appendix \ref{N>1}.
For the energy variance one finds in leading order in $\hbar$
\begin{eqnarray}
\left(\Delta{\cal H}\right)^2 & = & \frac{\hbar}{2}\sum_{a=1}^N\left[
\left\langle\frac{i}{\hbar}\left[\sqrt{\omega_a}q_a,{\cal H}\right]
\right\rangle^2
+\left\langle\frac{i}{\hbar}\left[\frac{p_a}{\sqrt{\omega_a}},{\cal H}\right]
\right\rangle^2\right]\nonumber\\
& & \qquad+{\cal O}\left(\hbar^2\right)
\label{oscergfluc5}\\
& = & \frac{\hbar}{2}\sum_{a=1}^N
\left[\omega_a\left\langle\partial_tq_a\right\rangle^2
+\frac{\left\langle\partial_tp_a\right\rangle^2}{\omega_a}\right]
\label{oscergfluc6}
\end{eqnarray}
with operator pairs $p_a$, $q_a$ and frequencies $\omega_a$, and the analog 
of Eq.~(\ref{oscergfluc3}) reads
\begin{equation}
\left(\Delta{\cal H}\right)^2
=\frac{\hbar}{2}\sum_{a=1}^N\left[\omega_a\left(\partial_t\xi_a\right)^2
+\frac{\left(\partial_t\pi_a\right)^2}{\omega_a}\right]
+{\cal O}\left(\hbar^2\right)\,.
\label{oscergfluc7}
\end{equation}

\subsection{Spin Systems}

\subsubsection{General Correlation Functions}

We consider again two arbitrary operators $A$, $B$ which are now functions
of a spin operator $\vec S$. The expectation value of the product $AB$ 
within an SU(2) coherent state $|z\rangle$ for spin length $S$ can be formulated
as
\begin{eqnarray}
\langle z|AB|z\rangle & = & \frac{2S+1}{\pi}\int
\frac{d^2w}{(1+|w|^2)^{2(S+1)}}\nonumber\\
& & \qquad\qquad\cdot
\langle z|e^{-w\tilde S^-/ \hbar}Ae^{w\tilde S^-/ \hbar}|z\rangle\nonumber\\
& & \qquad\qquad\cdot
\langle z|e^{\bar w\tilde S^+/ \hbar}Be^{-\bar w\tilde S^+/ \hbar}|z\rangle\,,
\end{eqnarray}
where we have used the completeness relation in the form (\ref{complsu24})
and the observation
\begin{equation}
e^{-\bar w\tilde S^+/ \hbar}|z\rangle=Ue^{-\bar wS^+/ \hbar}|0\rangle=|z\rangle
\end{equation}
with $U$ given in Eqs.~(\ref{su2U1})-(\ref{su2U3}). Employing now again the
expansion (\ref{itcom}) and performing the integration leads to
\begin{eqnarray}
\langle z|AB|z\rangle & = & \sum_{n=0}^{2S}\frac{(2S-n)!}{n!(2S)!}
\left\langle z\left|
\left[\frac{i}{\hbar}\tilde S^-,A\right]_n\right|z\right\rangle\nonumber\\
& & \qquad\qquad\cdot
\left\langle z\left|
\left[\frac{i}{\hbar}\tilde S^+,B\right]_n\right|z\right\rangle\,.
\label{su2exp1}
\end{eqnarray}
The above equation is the spin analog of the result (\ref{oscexp2}). Again all
iterated commutators are of the same order in $\hbar$ and $S$ whereas the
prefactor of the $n$-th term carries a product $2S(2S-1)\cdots(2S-n+1)$ in its
denominator. Thus, Eq.~(\ref{su2exp1}) is essentially an expansion in the
quantum parameter $1/S$. Note that the spin components 
$\tilde S^x$, $\tilde S^y$ represent the direction perpendicular to the
spin polarization of the coherent state $|z\rangle$. Alternatively,
the result (\ref{su2exp1}) can be written as
\begin{eqnarray}
\langle z|AB|z\rangle & = & \sum_{n=0}^{2S}\frac{(2S-n)!}{n!(2S)!}
\left\langle S\left|
\left[\frac{i}{\hbar}S^-,U^+AU\right]_n\right|S\right\rangle\nonumber\\
& & \qquad\qquad\cdot
\left\langle S\left|
\left[\frac{i}{\hbar}S^+,U^+BU\right]_n\right|S\right\rangle
\label{su2exp2}\\
& = & \sum_{n=0}^{2S}
\left\langle S\left|U^+AU\right|S-n\right\rangle\nonumber\\
& & \qquad\qquad\cdot
\left\langle S-n\left|U^+BU\right|S\right\rangle\,.
\label{su2exp2a}
\end{eqnarray}
Analogously to Eq.~(\ref{oscexp1b}) for
oscillatory systems, the last formulation is just the completeness relation
for the states $|m\rangle$ and allows for an alternative derivation
of the central result (\ref{su2exp1}).
Using the latter, 
arbitrary correlation functions within SU(2) coherent states can be
expressed in full analogy to Eq.~(\ref{oscexp3}).

\subsubsection{Fluctuations}

For the variance of an hermitian operator $A$ we have
\begin{equation}
\left(\Delta A\right)^2=\sum_{n=1}^{2S}\frac{(2S-n)!}{n!(2S)!}
\left|\left\langle z\left|
\left[\frac{i}{\hbar}\tilde S^-,A\right]_n\right|z\right\rangle\right|^2\,.
\label{su2exp3}
\end{equation}
The expectation values occurring in leading order can be rewritten as
\begin{eqnarray}
\left|\left\langle z\left|\left[i\tilde S^-,A\right]
\right|z\right\rangle\right|^2 & = &
\sum_{i=1}^3\left|\left\langle z\left|\left[i\tilde S^i,A\right]
\right|z\right\rangle\right|^2
\label{commid}\\
& = &\sum_{i=1}^3\left|\left\langle z\left|\left[iS^i,A\right]
\right|z\right\rangle\right|^2\,,
\end{eqnarray}
where we have observed that $|z\rangle$ is an eigenstate of $\tilde S^z$,
and that $\vec{\tilde S}$ and $\vec S$ are related by an orthogonal matrix,
\begin{equation}
\tilde S^i=\sum_{j=1}^3O_{ji}S^j\,.
\end{equation}
Thus, we have
\begin{equation}
\left(\Delta A\right)^2=\frac{1}{2S}
\sum_{i=1}^3\left|\left\langle z\left|\left[\frac{i}{\hbar}S^i,A\right]
\right|z\right\rangle\right|^2+{\cal O}\left(\frac{1}{S^2}\right)\,,
\label{su2exp4}
\end{equation}
and by a slight generalization of the above arguments one finds for the
expectation value of a product of commuting operators $A$, $B$
\begin{eqnarray}
& & \langle z|AB|z\rangle=\frac{1}{2}\langle z|AB+BA|z\rangle\nonumber\\
&  & \quad=\langle z|A|z\rangle\langle z|B|z\rangle\nonumber\\
&  & \qquad+\frac{1}{2S}
\sum_{i=1}^3\left\langle z\left|\left[\frac{i}{\hbar}S^i,A\right]
\right|z\right\rangle
\left\langle z\left|\left[\frac{i}{\hbar}S^i,B\right]
\right|z\right\rangle\nonumber\\
&  & \qquad+{\cal O}\left(\frac{1}{S^2}\right)\,.
\label{su2exp4a}
\end{eqnarray}
The requirement here for a symmetrized operator product stems from the
fact that for an identity analogous to Eq.~(\ref{commid}) to hold
products of expectation values involving both $\tilde S^x$ and $\tilde S^y$
should drop out.

Choosing now in Eq.~(\ref{su2exp4}) $A$ to be the Hamiltonian 
${\cal H}$ of the underlying system we
can write by the same arguments as for Eq.~(\ref{oscergfluc2})
\begin{equation}
\left(\Delta {\cal H}\right)^2=\frac{1}{2S}
\left\langle\partial_t\vec S\right\rangle^2+{\cal O}\left(\frac{1}{S^2}\right)
\label{su2ergfluc1}
\end{equation}
with $\langle\cdot\rangle=\langle z|\cdot|z\rangle$. To this result the same
comments apply as to its oscillatory counterpart Eq.~(\ref{oscergfluc2}):
If the system is initially in an SU(2) coherent state it holds
(cf. Eq.~(\ref{defsu2}))
\begin{equation}
\left\langle\partial_t\vec S\right\rangle=\hbar S\partial_t\vec s
\end{equation}
and
\begin{equation}
\left(\Delta {\cal H}\right)^2=\left(\hbar S\right)^2
\left(\frac{1}{2S}\left(\partial_t\vec s\right)^2
+{\cal O}\left(\frac{1}{S^2}\right)\right)
\label{su2ergfluc2}
\end{equation}
at initial time $t=t_i$, and for not too large times $t>t_i$ the
system will approximately remain coherent in its time evolution under ${\cal H}$
with $\vec s(t)$ being a classical vector. Our finding (\ref{su2ergfluc2}) is
again a manifestation of our previous result (\ref{oscergfluc3}):
In leading order in the quantum parameter ($\hbar$ or $1/S$) the variance
of the energy is due to the classical motion of the system.
Findings of the type (\ref{su2ergfluc2}) were also
obtained previously in Ref.~\cite{Schliemann98} 
on the example of specific Hamiltonians.
Here we provide a generalization to arbitrary systems based on the
very general expansions (\ref{su2exp1}),(\ref{su2exp3}) for
correlation functions and fluctuations.

Decoherence effects, i.e. deviations from the coherent state with
time-dependent parameters $\vec s(t)$ are again indicated by the higher-order
terms in the energy variance, as we shall investigate on a specific example
in section \ref{lmgiso}.  Conversely, the Zeeman Hamiltonian
(\ref{zeeman}) generates a strictly coherent time evolution with
\begin{equation}
\left(\Delta{\cal H}_z\right)^2
\equiv\frac{\left(\hbar S\right)^2}{2S}\left(\partial_t\vec s\right)^2
\label{su2ergfluc3}
\end{equation}
as an identity for arbitrary times $t\geq t_i$.

Similarly as for oscillator systems, the above results for general
operator products are easily generalized to the situation of $N>1$ spins
of various lengths; details can be found in appendix \ref{N>1}. The leading 
order of the energy variance is given by
\begin{equation}
\left(\Delta {\cal H}\right)^2=\sum_{a=1}^N\left[\frac{1}{2S_a}
\left\langle\partial_t\vec S_a\right\rangle^2
+{\cal O}\left(\frac{1}{S^2_a}\right)\right]\,,
\label{su2ergfluc4}
\end{equation}
and Eq.~(\ref{su2ergfluc2}) is generalized to
\begin{equation}
\left(\Delta {\cal H}\right)^2=\sum_{a=1}^N\left[\left(\hbar S_a\right)^2
\left(\frac{1}{2S_a}\left(\partial_t\vec s_a\right)^2
+{\cal O}\left(\frac{1}{S^2_a}\right)\right)\right]\,.
\label{su2ergfluc5}
\end{equation}

\section{The Lipkin-Meshkov-Glick model}
\label{lmg}

The Lipkin-Meshkov-Glick (LMG) model is an approximate description
of $N$ interacting spin-$1/2$ systems and was originally inspired by
nuclear physics \cite{Lipkin65,Meshkov65,Glick65}. More recently this model has
been argued to describe two-mode Bose-Einstein condensates 
\cite{Cirac98,Steel98,Zibold10,Gross10},
phase transitions in optical cavity QED \cite{Morrison08a,Morrison08b,Larson10},
and molecular magnets \cite{Campos11}.
Moreover it has been employed to model a spin bath \cite{Hamdouni07,Quan07}
and in studies of quenched dynamics \cite{Das06}. In the last decade a flurry
of publications investigating various aspects of the LMG model has appeared;
as an entry point to the recent literature we refer to 
Refs.\cite{Dusuel05,Leyvraz05,Unanyan05,Ribeiro08,Orus08,Caneva08,Ma09,Scherer09,Engelhardt13,Lerma13,Engelhardt14}.

Concentrating on the sector of maximal spin $S=N/2$, the LMG
Hamiltonian reads
\begin{equation}
{\cal H}=-hS^z-\frac{1}{2\hbar S}\left(\gamma_xS^xS^x+\gamma_yS^yS^y\right)\,,
\end{equation}
where $h$ can be interpreted as a magnetic field coupling to the
$z$-component of the spin while $\gamma_x$, $\gamma_y$ parametrize
an anisotropic interaction among the perpendicular components.
The factor $\hbar S$ in the denominator is a convention common to the
literature and leads to a linear scaling of energies as a function
of $S\gg 1$. The expectation value with in an SU(2) coherent state is given by
(neglecting a constant contribution)
\begin{eqnarray}
\langle{\cal H}\rangle & = & \hbar S\Bigl(-h\cos\vartheta\nonumber\\
 & & +\frac{\tilde\gamma_x}{2}\sin^2\vartheta\cos^2\varphi
+\frac{\tilde\gamma_y}{2}\sin^2\vartheta\sin^2\varphi\Bigr)
\end{eqnarray}
and equals the classical energy expression up the renormalized parameters
$\tilde\gamma_i=\gamma_i(1-1/(2S))$. Taking coherent expectation values
of both sides of the Heisenberg equations of motion one obtains the
(semi-)classical equations 
\begin{eqnarray}
\frac{ds^x}{dt} & = & h\sin\vartheta\sin\varphi
-\tilde\gamma_y\cos\vartheta\sin\vartheta\sin\varphi\,,
\label{lmgeom1}\\
\frac{ds^y}{dt} & = & -h\sin\vartheta\cos\varphi
+\tilde\gamma_x\cos\vartheta\sin\vartheta\cos\varphi\,,
\label{lmgeom2}\\
\frac{ds^z}{dt} & = & -\left(\tilde\gamma_x-\tilde\gamma_y\right)
\sin^2\vartheta\cos\varphi\sin\varphi\,.
\label{lmgeom3}
\end{eqnarray}
For the energy variance one finds by a direct (and somewhat tedious)
calculation
\begin{equation}
\left(\Delta{\cal H}\right)^2=\Omega_1+\Omega_2
\end{equation}
with
\begin{eqnarray}
& & \Omega_1=\left(\hbar S\right)^2\frac{1}{2S}\Bigl[
h^2\sin^2\vartheta-2h\tilde\gamma_x\cos\vartheta\sin^2\vartheta\cos^2\varphi
\nonumber\\
& & \quad-2h\tilde\gamma_y\cos\vartheta\sin^2\vartheta\sin^2\varphi\nonumber\\
& & \quad+\tilde\gamma_x^2\left(\sin^4\vartheta\cos^2\varphi\sin^2\varphi
+\cos^2\vartheta\sin^2\vartheta\sin^2\varphi\right)\nonumber\\
& & \quad+\tilde\gamma_y^2\left(\sin^4\vartheta\cos^2\varphi\sin^2\varphi
+\cos^2\vartheta\sin^2\vartheta\cos^2\varphi\right)\nonumber\\
& & \quad-2\tilde\gamma_x\tilde\gamma_y\sin^4\vartheta\cos^2\varphi\sin^2\varphi
\Bigr]
\end{eqnarray}
being of leading order $1/S$ while the contributions summarized in
\begin{eqnarray}
& & \Omega_2=\left(\hbar S\right)^2\frac{1}{8S^2}
\left(1-\frac{1}{2S}\right)\Bigl[-4\gamma_x\gamma_y\cos^2\vartheta\nonumber\\
& & +\left(\gamma_x\left(1-\sin^2\vartheta\cos^2\varphi\right)
+\gamma_y\left(1-\sin^2\vartheta\sin^2\varphi\right)\right)^2
\Bigr]\nonumber\\
\label{lmgomega2}
\end{eqnarray}
are of order $1/S^2$ and higher. Using now Eqs.~(\ref{lmgeom1})-(\ref{lmgeom3})
we can identify the leading contribution to the energy uncertainty as 
\begin{equation}
\Omega_1=\left(\hbar S\right)^2\frac{1}{2S}\left(\frac{d\vec s}{dt}\right)^2\,,
\end{equation}
in accordance with the general result (\ref{su2ergfluc2}). The
subleading contributions $\Omega_2$ indicate decoherence effects, i.e.
departures from the submanifold of the coherent states in the Hamiltonian
time evolution, as we now discuss explicitly on the example of the isotropic
LMG model.

\subsection{The isotropic case}
\label{lmgiso}

Putting $\gamma_x=\gamma_y=:\gamma$ the Hamiltonian becomes diagonal in the
states $|m\rangle$ with eigenvalues
\begin{equation}
\varepsilon_m/ \hbar=-hm+\frac{\gamma}{2S}m^2-\frac{\gamma}{2}(S+1)\,.
\end{equation}
This eigensystem is simple enough to analytically compute the exact
time evolution of coherent expectation values
$\langle\vec S(t)\rangle$: Due to symmetry, the $z$-component is constant, 
\begin{equation}
\langle S^z(t)\rangle\equiv\hbar S\cos\vartheta\,
\end{equation}
while for the perpendicular components one finds
\begin{eqnarray}
& & \langle S^+(t)\rangle=\hbar S\sin\vartheta
e^{i(\varphi-(h-\gamma(1-1/(2S))\cos\vartheta)t)}\nonumber\\
& & \cdot
\left(e^{-i\frac{\gamma\cos\vartheta}{2S} t}\left[\cos\left(\frac{\gamma}{2S}t\right)
+i\cos\vartheta\sin\left(\frac{\gamma}{2S}t\right)\right]\right)^{2S-1}\,.
\label{lmgS+1}
\end{eqnarray}
The above closed result relies on the fact that $S^+$ couples only
eigenstates with neighboring indices such that all occurring energy differences
are, apart from a constant term, linear in $m$. The first line in 
Eq.~(\ref{lmgS+1}) describes a classical rotation of the spin according to
Eqs.~(\ref{lmgeom1})-(\ref{lmgeom3}) whereas the second line contains
quantum effects: The ``spin length''
\begin{equation}
|\langle S^+(t)\rangle|=\hbar S\sin\vartheta
\left(1-\sin^2\vartheta\sin^2\left(\frac{\gamma}{2S}t\right)\right)^{S-1/2}
\end{equation}
composed from the perpendicular components breathes sinusoidally in time.
Quantum (quasi-)revivals occur at times at $t=2\pi kS/ \gamma$ for any
integer $k$ where the state returns precisely to the submanifold of the
coherent states. These times are large in the semiclassical regime as they
are proportional to $S$. 

Regarding small times, we define $t=:\sqrt{S}\tau$ and consider the regime
$\gamma\tau\ll\sqrt{S}$, such that for large $S\gg 1$ it follows
\begin{eqnarray}
\frac{|\langle S^+(t)\rangle|}{\hbar S\sin\vartheta} & \approx &
\left(1-\frac{(\gamma\tau\sin\vartheta)^2/4}{S}\right)^{S-1/2}\nonumber\\
& \approx & e^{-(\gamma\tau\sin\vartheta)^2/4}\,,
\end{eqnarray}
i.e. 
the spin expectation value $\langle\vec S(t)\rangle$ shows a gaussian decay
with time scale $\Delta t=\sqrt{2S}/(\gamma\sin\vartheta)$. 
On this time scale, sometimes known as Ehrenfest time \cite{Silvestrov02},
departures between classical and quantum dynamics become sizable.
The above finding for $\Delta t$
is consistent with a heuristic uncertainty argument in the following
sense: Replacing in $\Delta H\Delta t\geq\hbar$ the energy uncertainty
with $\sqrt{\Omega_2}$ one obtains a lower bound for $\Delta t$ being
proportional to $\hbar S$ which is a constant independent of S in the 
semiclassical regime. Thus, this lower bound is consistent with the above
result which grows with the square root of $S$.

\section{The Dicke Model}
\label{dicke}

The Dicke model describes the superradiant
interaction of a single cavity mode of a
radiation field with $N$ two level systems (atoms) \cite{Dicke54}.
Although introduced already in the 1950s, this model continues
to be investigated under various aspects; as a guide to the recent
literature see e.g. Refs.~\cite{Lambert04,Vidal06,Bastidas12,Bakemeier13}.

Focusing again on the sector of maximal spin $S=N/2$, the Dicke
Hamiltonian can be formulated as
\begin{eqnarray}
{\cal H} & = & \hbar\omega a^+a+\Omega S^z
+\frac{\lambda}{\sqrt{2S}}S^x\left(a^++a\right)\\
& = & \frac{1}{2}\left(p^2+\omega^2q^2\right)+\Omega S^z
+\lambda\sqrt{\frac{\omega}{\hbar S}}S^xq\,,
\end{eqnarray}
where the parameters $\omega$, $\Omega$, and $\lambda$ have all dimension
of inverse time.
In the classical limit, the superradiant phase, characterized by a finite
bosonic occupation in the ground state, occurs for $\lambda^2>\Omega\omega$.
The expectation value of the Hamiltonian within a tensor product of an
oscillator and an SU(2) coherent state reads
\begin{eqnarray}
\langle{\cal H}\rangle & = & \hbar\omega|\alpha|^2+\Omega\hbar S\cos\vartheta
\nonumber\\
& & +\frac{\lambda}{\sqrt{2S}}\hbar S\sin\vartheta\cos\varphi
\left(\bar\alpha+\alpha\right)\,,
\end{eqnarray}
which perfectly matches the classical expression. The (semi-)classical
equations of motion can be obtained analogously as 
Eqs.~(\ref{lmgeom1})-(\ref{lmgeom3}), 
\begin{eqnarray}
\frac{d\bar\alpha}{dt} & = & i\omega\bar\alpha
+\frac{i}{\hbar}\frac{\lambda}{\sqrt{2S}}\hbar S\sin\vartheta\cos\varphi\,,
\label{dickeeom1}\\
\frac{ds^x}{dt} & = & -\Omega\sin\vartheta\sin\varphi\,,
\label{dickeeom2}\\
\frac{ds^y}{dt} & = & \Omega\sin\vartheta\cos\varphi
-\frac{\lambda}{\sqrt{2S}}\cos\vartheta
\left(\bar\alpha+\alpha\right)\,,
\label{dickeeom3}\\
\frac{ds^z}{dt} & = & \frac{\lambda}{\sqrt{2S}}\sin\vartheta\sin\varphi
\left(\bar\alpha+\alpha\right)\,,
\label{dickeeom4}
\end{eqnarray}
and a direct (but again quite lengthy) calculation of the
energy variance yields
\begin{equation}
\left(\Delta{\cal H}\right)^2=\Omega_1+\Omega_2
\end{equation}
with the leading-order term
\begin{eqnarray}
& & \Omega_1=(\hbar\omega)^2|\alpha|^2
+\frac{(\hbar S)^2}{2S}\Omega^2\sin^2\vartheta\nonumber\\
& & \quad+(\hbar S)^2\frac{\lambda^2}{2S}\Biggl[\frac{1}{2S}
\left(\sin^2\vartheta\sin^2\varphi+\cos^2\vartheta\right)
\left(\bar\alpha+\alpha\right)^2\nonumber\\
& & \qquad\qquad\qquad+\sin^2\vartheta\cos^2\varphi\Biggr]\nonumber\\
& & \quad+\hbar\omega\frac{\lambda}{\sqrt{2S}}\hbar S\sin\vartheta\sin\varphi
\left(\bar\alpha+\alpha\right)\nonumber\\
& & \quad-\Omega\frac{\lambda}{\sqrt{2S}}\frac{(\hbar S)^2}{S}
\cos\vartheta\sin\vartheta\cos\varphi\left(\bar\alpha+\alpha\right)
\end{eqnarray}
and the subleading contributions
\begin{eqnarray}
\Omega_2 & = & (\hbar S)^2\frac{\lambda^2}{(2S)^2}
\left(\sin^2\vartheta\sin^2\varphi+\cos^2\vartheta\right)\\
& = & \frac{\lambda^2}{(2S)^2}
\left(\left\langle S^y\right\rangle^2+\left\langle S^z\right\rangle^2\right)\,.
\label{dickeomega2}
\end{eqnarray}
Finally, comparison with Eqs.~(\ref{dickeeom1})-(\ref{dickeeom4}) shows
\begin{equation}
\Omega_1=\frac{\hbar}{2}\left(\omega\left(\partial_t\xi\right)^2
+\frac{\left(\partial_t\pi\right)^2}{\omega}\right)
+\frac{\left(\hbar S\right)^2}{2S}\left(\partial_t\vec s\right)^2\,,
\end{equation}
in accordance with the general results (\ref{oscergfluc7}), (\ref{su2ergfluc5}),
and (\ref{combfluc}).

Note that the higher-order contributions (\ref{dickeomega2}) describing
decoherence effects depend only on the coupling parameter $\lambda$ but not
on the frequencies $\omega$, $\Omega$, in accordance with the fact that
spin and oscillator show perfectly coherent time evolutions in the absence
of coupling. Another distinctive feature of the result (\ref{dickeomega2})
(compared to e.g. Eq.~(\ref{lmgomega2})) is its simplicity which calls for
further applications. In fact, an extensive numerical study of the dynamics
of the Dicke model in the semiclassical regime was performed recently
in Ref.~\cite{Bakemeier13}. Here the initial condition was, as in the present
work, a tensor product of an an oscillator and a spin coherent state, and
it is straightforward to evaluate
the above $\Omega_2$ in terms of such dynamical data. In particular, it is an
interesting speculation whether or not  $\Omega_2$ behaves differently in
the regular versus (quantum) chaotic regime as studied in 
Ref.~\cite{Bakemeier13}.
Another aspect is to compare the Ehrenfest times $\Delta t$
found numerically with
estimates according to $\sqrt{\Omega_2}\Delta t\geq\hbar$.

\section{Coherent Intertwiners in Spin Networks}
\label{intertw}

We now apply our general findings on coherent expectation values of operator
products to spin network states as studied in Loop Quantum Gravity (LQG)
\cite{Rovelli14,Rovelli11,Perez13}. In brief, a spin network is a 
collection of points (called {\em vertices} or {\em nodes})
in (typically) three-dimensional space
connected by one-dimensional curves ({\em edges}). Each edge is assigned a 
spin of individual length, and a spin network state in the tensor product
of all those SU(2) representations is defined by the additional requirement
that all spins joining in a given node are coupled to a total singlet.
The latter property implements the Gauss constraint on the holonomy and
flux variables used in LQG \cite{Rovelli04,Thiemann07}.

A convenient parametrization of spin network states are coherent intertwiners
as introduced by Livine and Speziale \cite{Livine07}.
Fixing an $N$-valent node (connecting $N$ edges), one considers a
tensor product
\begin{equation}
|\Phi\rangle:=\bigotimes_{a=1}^N|\vartheta_a,\varphi_a\rangle
\label{nodecostate}
\end{equation}
of SU(2) coherent states describing the spin on each edge.
A coherent intertwiner is then defined by the  
projection of this object onto the singlet subspace \cite{Livine07}
\begin{equation}
|\Phi\rangle_s=\frac{P|\Phi\rangle}
{\sqrt{\langle\Phi|P|\Phi\rangle}}\,,
\label{defcoint}
\end{equation}
where the denominator takes care of the normalization. The projection
operator can be formalized by a Haar integration over all uniform
rotations of the $N$ spins (group averaging),
\begin{eqnarray}
P & = & \int_{\rm SU(2)}d\mu\exp\left(i\psi\vec n\sum_a\vec S_a\right)
\label{defP1}\\
& = & \frac{1}{4\pi^2}\int_0^{\pi}d\vartheta\sin\vartheta\int_0^{2\pi}d\varphi
\int_0^{2\pi}d\psi\sin^2\frac{\psi}{2}\nonumber\\
& & \qquad\cdot\exp\left(i\psi\vec n\sum_a\vec S_a\right)
\label{defP2}
\end{eqnarray}
with $\vec n=(\sin\vartheta\cos\varphi,\sin\vartheta\sin\varphi,\cos\vartheta)$.
Here and in what follows we take all spin operators to be dimensionless 
(as a factor of $\hbar$ will occur below in the Planck length squared).
In particular, a coherent intertwiner is by construction invariant under
arbitrary rotations of all spins meaning
\begin{equation}
\left(\sum_{a=1}^N\vec S_a\right)|\Phi\rangle_s=0\,.
\label{clorelquant}
\end{equation}

Moreover, nodes in a spin network allow for a geometric interpretation in terms
of convex polyhedra \cite{Bianchi11a}. From a classical point of view,
this relies on a theorem due to Minkowski \cite{Minkowski1897}. It states that
given $N$ unit vectors $\vec s_a$ and $N$ positive numbers $A_a$ 
fulfilling $\sum_aA_a\vec s_a=0$, there is
a unique convex polyhedron with $N$ faces such that $\vec s_a$ is the normal
to the $a$-th face and $A_a$ is its area. Thus choosing as areas the
quantum numbers $S_a$, the classical closure relation
\begin{equation}
\sum_{a=1}^NS_a\vec s_a=0
\label{clorelclass}
\end{equation}
ensures that the geometric information contained in the
state (\ref{nodecostate}) encodes a convex polyhedron. The quantum
counterpart of the relation (\ref{clorelclass}) is equation (\ref{clorelquant})
giving rise to the notion of a {\em quantum polyhedron} \cite{Bianchi11a}.
In the framework of LQG, the spin operators representing the faces of the
polyhedron are, up to a prefactor, considered to be flux operators
\cite{Rovelli04,Thiemann07} 
\begin{equation}
\vec E_a=8\pi\gamma\ell_P^2\vec S_a
\end{equation}
with $\gamma$ being the Immirzi parameter and the squared
Planck length $\ell_P^2=\hbar G/c^3$. 

Let us now explore expectation values within coherent intertwiners. Here one
can concentrate without loss of generality  on operators unchanged by 
uniform rotations since for any operator being the sum of a rotationally
invariant part and terms without this property, only the former will
contribute.
Any rotationally invariant operator $Q$ commutes
with the projector onto the singlet space, $[Q,P]=0$, such that $PQP=QP=PQ$.
Therefore we can use the result (\ref{su2exp4a}) to obtain a semiclassical
approximation to the expectation value within a coherent intertwiner,
\begin{eqnarray}
& & _s\langle\Phi|Q|\Phi\rangle_s
=\frac{\langle\Phi|PQ+QP|\Phi\rangle}
{2\left\langle\Phi\left|P\right|\Phi\right\rangle}\\ 
& & \quad=\langle\Phi|Q|\Phi\rangle\nonumber\nonumber\\
& & \qquad+\sum_{a=1}^N\frac{1}{2S_a}\sum_{i=1}^3
\left\langle\Phi\left|\left[iS_a^i,Q\right]\right|\Phi\right\rangle
C_a^i(\Phi)+\cdots\nonumber\\
\label{expQ}
\end{eqnarray}
with
\begin{equation}
C_a^i(\Phi)=
\frac{\left\langle\Phi\left|\left[iS_a^i,P\right]\right|\Phi\right\rangle}
{\left\langle\Phi\left|P\right|\Phi\right\rangle}\,.
\label{defC}
\end{equation}
Thus, the expectation value of $Q$ is in leading order
just given by the expectation value of the unprojected state 
(\ref{nodecostate}), and the normalization factor in the definition
(\ref{defcoint}) drops out. For the subleading corrections one needs to
determine the coefficients (\ref{defC}). Here both numerator and
denominator are conveniently formulated in terms of Haar integrations as
shown explicitly in Eq.~(\ref{defP2}). In the semiclassical regime studied
here where all spins are long, $\forall a S_a\gg 1$, these integrals become
amenable to a saddle-point approximation as worked out in Ref.~\cite{Livine07}.
For the denominator one finds for a general $N$-valent node 
\begin{equation}
\left\langle\Phi\left|P\right|\Phi\right\rangle
=\frac{1}{\sqrt{\pi\det H}}
+\frac{{\rm tr}\left(H^{-1}\right)}{4\sqrt{\pi\det H}}+\cdots
\label{ctnormres}
\end{equation}
where
\begin{equation}
H^{ij}=\sum_{a=1}^NS_a\left(\delta^{ij}-s_a^is_a^j\right)
\label{defHmat}
\end{equation}
is twice the negative Hessian of the saddle point expression, 
and the details of the calculation can be found in appendix \ref{sadpo}.
Since $H$ is a linear combination of geometric projection operators
$(\delta^{ij}-s_a^is_a^j)$ with positive coefficients, its eigenvalues are
nonnegative, and zero eigenvalues only occur in the degenerate case where
all vectors $\vec s_a$ are collinear, which we shall not consider here.
Thus, the eigenvalues of $H$ can be taken to be positive, and the determinant
can be formulated more explicitly as \cite{Livine07} 
\begin{eqnarray}
\det H & = & \frac{T}{2}\sum_{ab}S_aS_b\left(\vec s_a\times\vec s_b\right)^2
\nonumber\\
& & -\frac{1}{6}\sum_{abc}S_aS_bS_c
\left|\left(\vec s_a\times\vec s_b\right)\cdot\vec s_c\right|^2
\end{eqnarray}
with $T=\sum_aS_a$. The semiclassical limit of a quantum polyhedron
is obtained by rescaling all quantum numbers as $S_a\mapsto \lambda S_a$
with some integer $\lambda\gg 1$. Thus the leading term in Eq.~(\ref{ctnormres})
(already obtained in Ref.~\cite{Livine07})
is of order $\lambda^{-3/2}$ while the subleading correction scales
like $\lambda^{-5/2}$. The numerator in Eq.~(\ref{defC})
can be evaluated via saddle point approximation in a similar fashion
(see appendix \ref{sadpo}) giving, again for a general $N$-valent node
fulfilling the classical closure relation (\ref{clorelclass}), 
\begin{equation}
\left\langle\Phi\left|\left[i\vec S_a,P\right]
\right|\Phi\right\rangle
=S_a\frac{\vec s_a\times\left(H^{-1}\vec s_a\right)}{\sqrt{\pi\det H}}
\label{numeratorres}
\end{equation}
such that for the coefficients themselves we have the amazingly simple
result
\begin{equation}
\vec C_a(\Phi)=S_a\vec s_a\times\left(H^{-1}\vec s_a\right)\,.
\label{Cres}
\end{equation}
The expression (\ref{numeratorres}) is of order
$\lambda^{-3/2}$ while
the coefficients (\ref{Cres}) are independent of $\lambda$ and vanish if the
matrix $H$ is proportional to the unit matrix. Thus,
polyhedra where all eigenvalues of $H$ are degenerate enjoy an enhanced
classical character in the sense that the leading order of 
semiclassical corrections to
general expectation values (\ref{expQ}) vanishes.
In the general case, Eq.~(\ref{expQ})
tells us that the coherent-intertwiner expectation value of any (rotationally
invariant) operator is in leading order given by the expectation value
of the unprojected tensor product of SU(2) coherent states, and the 
leading correction scales with the inverse of the spin lengths.

The  coefficients (\ref{Cres}) are universal in the sense that they are
the same for any operator $Q$.
Making use of the symmetry of $H$ they can also be formulated as
\begin{equation}
C_a^i(\Phi)=S_a\sum_{jkl}\epsilon^{ijk}
\left(H^{-1}\right)^{kl}\left(s_a^ls_a^j-\delta^{lj}\right)
\label{srule1}
\end{equation}
implying the sum rule
\begin{equation}
\sum_{a=1}^NC_a^i(\Phi)=-\sum_{jkl}\epsilon^{ijk}
\left(H^{-1}\right)^{kl}H^{lj}=0\,,
\label{srule2}
\end{equation}
which also follows from the definition (\ref{defC}) and the quantum closure
relation (\ref{clorelquant}). Thus, the sum rule (\ref{srule2}) holds
independently of the fulfillment of the classical closure relation
(\ref{clorelclass}) which underlies the explicit result (\ref{Cres}).

Moreover, it is interesting to 
note that the matrix $H$ can be interpreted as the inertia tensor
of a distribution of masses $S_a$ whose positions are given by the unit
vectors $\vec s_a$. By the same token, the classical closure
relation (\ref{clorelclass}) states that the center of mass of this distribution
lies in the origin of the chosen coordinate system. 
In particular, $H$ is proportional to the unit matrix (such that the
expansion coefficients (\ref{defC}) vanish) if the node has the shape
of an archimedian body such as a regular tetrahedron.   
We leave it to further
studies to explore further possible consequences of the above analogy.

Very typical examples of rotationally invariant operators are volume
operators of polyhedra \cite{Rovelli95,Ashtekar95,Bianchi11a}.
The simplest nontrivial case of a quantum
polyhedron is given by a tetrahedron, 
i.e. a $4$-valent node \cite{Barbieri98}. 
The volume operator can be formulated as
\begin{equation}
V=\frac{\sqrt{2}}{3}\sqrt{|\vec E_1\cdot(\vec E_2\times\vec E_3)|}
\label{defV}
\end{equation}
using any three of the four flux operators. Squaring this expressions and
striping all prefactors one is led to consider the expression
\begin{equation}
Q=\vec S_1\cdot(\vec S_2\times\vec S_3)\,.
\end{equation}
acting on the Hilbert space defined by the constraint (\ref{clorelquant}).
The study of this operator in the semiclassical limit has attracted
quite a deal of interest recently 
\cite{Bianchi11a,Bianchi11b,Bianchi12,Schliemann13,Schliemann14}.
For the expectation value within coherent intertwiners one finds from 
Eqs.~(\ref{expQ}),(\ref{Cres})
\begin{eqnarray}
& & _s\langle\Phi|\vec S_1(\vec S_2\times\vec S_3)|\Phi\rangle_s
=S_1S_2S_3\Biggl[\vec s_1\left(\vec s_2\times\vec s_3\right)\nonumber\\
& & \qquad+\frac{1}{2}\Bigl(\vec s_1\times(\vec s_2\times\vec s_3)
\cdot\left(\vec s_1\times\left(H^{-1}\vec s_1\right)\right)
+{\rm c.p.}\Bigr)\Biggr]\nonumber\\
& & \qquad+\cdots
\label{expVsq}
\end{eqnarray}
As before, the form of the subleading corrections here holds if the classical 
closure relation (\ref{clorelclass}) is fulfilled. 

\section{Summary and Outlook}
\label{concl}

We have derived
general systematic expansions with respect to quantum parameters of
expectation values of products of arbitrary operators within both
oscillator coherent states and SU(2) coherent states.
These results are a versatile tools for the
study of the semiclassical regime of generic quantum systems.
In particular, we prove that the energy fluctuations of an 
arbitrary Hamiltonian are in leading order entirely due to the 
time dependence of the classical variables, a result very general and
very intuitive at the same time.

Our findings offer many possibilities for application in various fields
of physics. Here we have specifically studied the Dicke model stemming
from quantum optics, and the LMG model originating from nuclear physics.
For the latter system we have investigated decoherence effects (i.e. deviations
from the submanifold of coherent states) via an exact solution of the
dynamics, which also appears novel to the literature.
Finally we have applied our general results to coherent intertwiners 
in spin networks as investigated
in LQG. For expectation values of rotationally invariant operators (and these
are the only ones contributing) one finds here a subleading correction
to the classical limit given in terms of universal (i.e. operator-independent)
expansion coefficients which contain only geometric information about
the network node.

\appendix

\vspace{2cm}

\section{$N>1$ degrees of freedom}
\label{N>1}

Let us now extend our results on the coherent-state expectation values of
operator products to systems with $N>1$ degrees of freedom. We start
by two oscillatory degrees of freedom $q_a$, $p_a$ with frequencies
$\omega_a$, $a\in\{1,2\}$. Iterating the arguments leading to
Eq.~(\ref{oscexp2}) one finds
\begin{eqnarray}
\langle\alpha|AB|\alpha\rangle & = & \sum_{m,n=0}^{\infty}
\frac{\hbar^{m+n}}{m!n!}
\left\langle\alpha\left|\left[Q_2,\left[Q_1,A\right]_n\right]_m
\right|\alpha\right\rangle\nonumber\\
& & \cdot\left\langle\alpha\left|\left[Q_2^+,\left[Q_1^+,B\right]_n\right]_m
\right|\alpha\right\rangle\,.
\label{apposcexp1}
\end{eqnarray}
with $|\alpha\rangle=|\alpha_1\rangle\otimes|\alpha_2\rangle$ and
\begin{equation}
Q_a=\frac{i}{\sqrt{2}\hbar}\left(\sqrt{\omega_a}q_a
-i\frac{p_a}{\sqrt{\omega_a}}\right)\,.
\end{equation}
Since $Q_1$, $Q_2$ commute, the corresponding
left arguments in the above nested commutators
can be freely interchanged such that
\begin{widetext}
\begin{eqnarray}
& & \frac{(m+n)!}{m!n!}
\left\langle\alpha\left|\left[Q_2,\left[Q_1,A\right]_n\right]_m
\right|\alpha\right\rangle
\left\langle\alpha\left|\left[Q_2^+,\left[Q_1^+,B\right]_n\right]_m
\right|\alpha\right\rangle\nonumber\\
& & \qquad=\sum_{P_{mn}}\left\langle\alpha\left|
\left[Q_{P_{mn}(1)},\left[Q_{P_{mn}(2)},\cdots\left[Q_{P_{mn}(m+n)},A\right]
\cdots\right]\right]\right|\alpha\right\rangle\nonumber\\
& & \qquad\qquad\cdot\left\langle\alpha\left|
\left[Q_{P_{mn}(1)}^+,\left[Q_{P_{mn}(2)}^+,\cdots\left[Q_{P_{mn}(m+n)}^+,B\right]
\cdots\right]\right]\right|\alpha\right\rangle\,,
\end{eqnarray}
\end{widetext}
where the sum goes over all functions 
$P_{mn}:\{1,\dots,m+n\}\rightarrow\{1,2\}$ taking $m$ times the value $2$ and
$n$ times the value $1$. Thus we arrive at
\begin{widetext}
\begin{eqnarray}
\langle\alpha|AB|\alpha\rangle & = & \sum_{n=0}^{\infty}
\frac{\hbar^n}{n!}\sum_{P_n}
\left\langle\alpha\left|
\left[Q_{P_n(1)},\left[Q_{P_n(2)},\cdots\left[Q_{P_n(n)},A\right]
\cdots\right]\right]\right|\alpha\right\rangle\nonumber\\
& & \qquad\qquad\cdot\left\langle\alpha\left|
\left[Q_{P_n(1)}^+,\left[Q_{P_n(2)}^+,\cdots\left[Q_{P_n(n)}^+,B\right]
\cdots\right]\right]\right|\alpha\right\rangle\,,
\label{apposcexp2}
\end{eqnarray}
\end{widetext}
where the second sum extends now over all functions 
$P_n:\{1,\dots,n\}\rightarrow\{1,2\}$. Moreover, it is straightforward to see
that the above expression also holds for an arbitrary number $N$ of
oscillatory degrees of freedom with 
$|\alpha\rangle=|\alpha_1\rangle\otimes\cdots\otimes|\alpha_N\rangle$
and functions $P_n:\{1,\dots,n\}\rightarrow\{1,\dots,N\}$. In particular,
the variance of an hermitian operator $A$ can be expressed as
\begin{widetext}
\begin{equation}
\left(\Delta A\right)^2=\sum_{n=1}^{\infty}
\frac{\hbar^n}{n!}\sum_{P_n}\left|
\left\langle\alpha\left|
\left[Q_{P_n(1)},\left[Q_{P_n(2)},\cdots\left[Q_{P_n(n)},A\right]
\cdots\right]\right]\right|\alpha\right\rangle\right|^2\,,
\label{apposcexp3}
\end{equation}
\end{widetext}
and the leading-order results for energy fluctuations are given
in Eqs.~(\ref{oscergfluc5})-(\ref{oscergfluc7}).

The counterpart of Eq.~(\ref{apposcexp1}) for two spins $\vec S_1$, $\vec S_2$
reads
\begin{widetext}
\begin{equation}
\langle z|AB|z\rangle=\sum_{m=0}^{2S_2}\sum_{n=0}^{2S_1}
\frac{(2S_2-m)!}{m!(2S_2)!}\frac{(2S_1-n)!}{n!(2S_1)!}
\left\langle z\left|
\left[\frac{i}{\hbar}\tilde S_2^-,
\left[\frac{i}{\hbar}\tilde S_1^-,A\right]_n\right]_m
\right|z\right\rangle
\left\langle z\left|
\left[\frac{i}{\hbar}\tilde S_2^+,
\left[\frac{i}{\hbar}\tilde S_1^+,B\right]_n\right]_m
\right|z\right\rangle\,.
\label{appsu2exp1}
\end{equation}
\end{widetext}
with $|z\rangle=|z_1\rangle\otimes|z_2\rangle$.
Due to the more complicated prefactors, a similarly compact form
as in Eq.~(\ref{apposcexp2}) for the full expansion
seems to be unachievable for spin systems.
The leading terms of energy fluctuations given in 
Eqs.~(\ref{su2ergfluc4}),(\ref{su2ergfluc5}), however, are again rather
simple and allow for an intuitive interpretation.

Finally, combining both types of systems, the leading-order
contribution to the fluctuation of an Hamiltonian depending on $N$
oscillatory degrees of freedom and $M$ spins reads
\begin{eqnarray}
\left(\Delta{\cal H}\right)^2 & = & \frac{\hbar}{2}\sum_{a=1}^N
\left[\omega_a\left\langle\partial_tq_a\right\rangle^2
+\frac{\left\langle\partial_tp_a\right\rangle^2}{\omega_a}\right]
+{\cal O}\left(\hbar^2\right)\nonumber\\
& & +\sum_{b=1}^M\left[\left(\hbar S_b\right)^2
\left(\frac{1}{2S_b}\left(\partial_t\vec s_b\right)^2
+{\cal O}\left(\frac{1}{S^2_b}\right)\right)\right]\,.\nonumber\\
\label{combfluc}
\end{eqnarray}

\section{The Normalization of Coherent Intertwiners and Related Integrals}
\label{sadpo}

In order to evaluate the normalization integral of coherent intertwiners
in the semiclassical regime, we shall use a slightly different version
of SU(2) coherent states generated by
\begin{equation}
V(\vartheta,\varphi)=e^{-\varphi S^z}e^{-\vartheta S^y}
\label{su2V}
\end{equation}
such that compared to Eq.~(\ref{su2cos1}) one has 
(dropping again factors of $\hbar$)
\begin{equation}
V(\vartheta,\varphi)|S\rangle=e^{i\varphi S}U(\vartheta,\varphi)|S\rangle\,,
\end{equation} 
i.e. the coherent states generated by the operators (\ref{su2U1}) and
(\ref{su2V}) just differ by a phase factor which drops out from all 
expectation values. The operator (\ref{su2V}) fulfills
\begin{equation}
V^+\vec SV=\vec uS^x+\vec vS^y+\vec sS^z
\label{Vtrans}
\end{equation}
with
\begin{equation}
\vec u=\frac{(\vec e_z\times\vec s)\times\vec s}{|\vec e_z\times\vec s|}
\quad,\quad 
\vec v=\frac{\vec e_z\times\vec s}{|\vec e_z\times\vec s|}
\end{equation}
such that $\vec u$,  $\vec v$,  $\vec s$, form an orthonormal system.
If one had used the original operator (\ref{su2U1}) for generating
coherent states the form of the vectors $\vec u$,  $\vec v$ 
would be less transparent. Now the normalization integral can be written as
\begin{equation}
\langle\Phi|P|\Phi\rangle=\int d\mu\prod_{a=1}^N
\left\langle S_a\left|e^{i\psi\vec n_a\vec S_a}\right|S_a\right\rangle
\label{ctnorm1}
\end{equation}
with
\begin{equation}
n_a^x=\vec n\vec u_a\quad,\quad
n_a^y=\vec n\vec v_a\quad,\quad
n_a^z=\vec n\vec s_a
\label{defna}
\end{equation}
where $\vec n$ is the rotation axis occurring in 
Eqs.~(\ref{defP1}),(\ref{defP2}).
Taking into account the
explicit form of the rotation matrix element \cite{Edmonds57}
\begin{equation}
\left\langle S\left|e^{i\psi\vec n\vec S}\right|S\right\rangle
=\left(\cos\frac{\psi}{2}+in^z\sin\frac{\psi}{2}\right)^{2S}
\end{equation}
elementary manipulations lead to 
\begin{eqnarray}
\langle\Phi|P|\Phi\rangle & = & \frac{1}{2\pi^2}
\int_0^{\pi}d\vartheta\sin\vartheta\int_0^{2\pi}d\varphi
\int_0^{\pi/2}d\psi\sin^2\psi\nonumber\\
& & \quad\cdot\sum_{\eta=\pm}\prod_{a=1}^N
\left(\eta\cos\psi+i\vec n\vec s_a\sin\psi\right)^{2S_a}\,,
\label{ctnorm2}
\end{eqnarray}
where the cosine of $\psi$ is nonnegative in the entire integration interval.
Following Ref.~\cite{Livine07} we introduce 
$\vec p:=\vec n\sin\psi$ fulfilling
\begin{equation}
d^3p=\sin\vartheta\sin^2\psi\cos\psi d\vartheta d\varphi d\psi
\end{equation}
such that 
\begin{equation}
\langle\Phi|P|\Phi\rangle=\frac{1}{2\pi^2}\sum_{\eta=\pm}
\int_{p\leq 1}\frac{d^3p}{\sqrt{1-p^2}}e^{S_{\eta}(\vec p)}
\label{ctnorm3}
\end{equation}
where
\begin{equation}
S_{\eta}(\vec p)=\sum_{a=1}^N2S_a\ln\left(\eta\sqrt{1-p^2}+i\vec p\vec s_a\right)
\,.
\end{equation}
In this form the integral can be evaluated via saddle point
approximation to $S_{\eta}(\vec p)$. As discussed in detail in 
Ref.~\cite{Livine07}, provided that the classical closure relation
(\ref{clorelclass}) holds, the maximum of $S_{\eta}(\vec p)$ occurs at
$\vec p=0$ with 
\begin{equation}
S_+(0)=0\quad,\quad S_-(0)=2\pi i\sum_{a=1}^NS_a\,
\end{equation}
and since the latter sum must be integer for a nontrivial singlet space we
have $\exp(S_{\pm}(0))$=1. The Hessian is given by (cf. Eq.~(\ref{defHmat}))
\begin{equation}
\left(\frac{\partial^2S_{\pm}(\vec p)}
{\partial p^i\partial p^j}\right)_{\vec p=0}
=:-2H^{ij}=-2\sum_{a=1}^NS_a\left(\delta^{ij}-s_a^is_a^j\right)\,.
\label{hessian}
\end{equation}
Extending now the integration domain in Eq.~(\ref{ctnorm3}) to the
infinite space (as the integrand falls off rapidly), we are left with simple
gaussian integrals leading to the result (\ref{ctnormres}) where the
leading first term was already obtained in Ref.~\cite{Livine07} while the
subleading correction stems from expanding the square root in 
Eq.~(\ref{ctnorm3}).

To compute the numerator of the coefficients (\ref{defC}) we consider
\begin{eqnarray}
& & \left\langle\Phi\left|\left[iV_a\vec S_aV_a^+,P\right]
\right|\Phi\right\rangle
=\int d\mu\left\langle S_a\left|\left[i\vec S_a,e^{i\psi\vec n_a\vec S_a}\right]
\right|S_a\right\rangle\nonumber\\
& & \qquad\qquad\qquad\qquad\qquad\cdot
\prod_{b\neq a}\left\langle S_b\left|e^{i\psi\vec n_b\vec S_b}\right|S_b\right\rangle
\label{coeff1}
\end{eqnarray}
with $V_a=V(\vartheta_a,\varphi_a)$. With the help of the rotation 
matrix element \cite{Edmonds57}
\begin{eqnarray}
\left\langle S-1\left|e^{i\psi\vec n\vec S}\right|S\right\rangle & = &
\sqrt{2S}\left(\cos\frac{\psi}{2}+in^z\sin\frac{\psi}{2}\right)^{2S-1}\nonumber\\
& & \qquad\cdot(n^x+in^y)\sin\frac{\psi}{2}
\end{eqnarray}
one derives
\begin{eqnarray}
\left\langle S_a\left|\left[iS_a^x,e^{i\psi\vec n_a\vec S_a}\right]
\right|S_a\right\rangle 
& = & -i\left(\cos\frac{\psi}{2}+in_a^z\sin\frac{\psi}{2}\right)^{2S-1}
\nonumber\\
& & \cdot2S_an_a^y\sin\frac{\psi}{2}\,,\\
\left\langle S_a\left|\left[iS_a^y,e^{i\psi\vec n_a\vec S_a}\right]
\right|S_a\right\rangle 
& = & i\left(\cos\frac{\psi}{2}+in_a^z\sin\frac{\psi}{2}\right)^{2S-1}
\nonumber\\
& & \cdot2S_an_a^x\sin\frac{\psi}{2}\,,\\
\left\langle S_a\left|\left[iS_a^z,e^{i\psi\vec n_a\vec S_a}\right]
\right|S_a\right\rangle & = & 0\,.
\end{eqnarray}
Now proceeding as before the two nontrivial expectation values can be
formulated as
\begin{eqnarray}
& & \left\langle\Phi\left|\left[iV_aS_a^xV_a^+,P\right]
\right|\Phi\right\rangle\nonumber\\
& & =\frac{-iS_a}{\pi^2}\sum_{\eta=\pm}
\int\frac{d^3p}{\sqrt{1-p^2}}
\frac{\vec p\vec v_a}{\eta\sqrt{1-p^2}+i\vec p\vec s_a}
e^{S_{\eta}(\vec p)}\,,\nonumber\\\\
& & \left\langle\Phi\left|\left[iV_aS_a^yV_a^+,P\right]
\right|\Phi\right\rangle\nonumber\nonumber\\
& & =\frac{iS_a}{\pi^2}\sum_{\eta=\pm}
\int\frac{d^3p}{\sqrt{1-p^2}}
\frac{\vec p\vec u_a}{\eta\sqrt{1-p^2}+i\vec p\vec s_a}
e^{S_{\eta}(\vec p)}\,.\nonumber\\
\end{eqnarray}
Performing again a saddle point approximation to the exponential and
expanding the remaining integrand in quadratic order around $\vec p=0$ 
leads to 
\begin{eqnarray}
\left\langle\Phi\left|\left[iV_aS_a^xV_a^+,P\right]
\right|\Phi\right\rangle & = &
-S_a\frac{\left(\vec v_a^TH^{-1}\vec s_a\right)}
{\sqrt{\pi\det H}}\,,
\nonumber\\\\
\left\langle\Phi\left|\left[iV_aS_a^yV_a^+,P\right]
\right|\Phi\right\rangle & = &
S_a\frac{\left(\vec u_a^TH^{-1}\vec s_a\right)}
{\sqrt{\pi\det H}}\,,
\nonumber\\
\end{eqnarray}
and using Eq.~(\ref{Vtrans}) along with elementary geometric relations
it follows for the coefficients (\ref{defC})
\begin{eqnarray}
\vec C_a(\Phi) & = & -S_a
\left(\vec u_a\left(\vec v_a^TH^{-1}\vec s_a\right)
-\vec v_a\left(\vec u_a^TH^{-1}\vec s_a\right)\right)\nonumber\\
& = & S_a\vec s_a\times
\Bigl(\vec v_a\left(\vec v_a^TH^{-1}\vec s_a\right)\nonumber\\
& & \qquad\qquad\qquad\quad
+\vec u_a\left(\vec u_a^TH^{-1}\vec s_a\right)\Bigr)\,.
\end{eqnarray}
Finally, observing that $\vec u_a$, $\vec v_a$ span the plane
perpendicular to $\vec s_a$ we obtain the result (\ref{Cres}), and the
numerator of Eq.~(\ref{defC}) is given by Eq.~(\ref{numeratorres}).

\end{document}